\documentclass[aps,twocolumn,showpacs,preprintnumbers,nofootinbib,float,longbibliography,superscriptaddress]{revtex4-2}

\usepackage{epsfig}
\usepackage{graphicx}
\usepackage{palatino}
\usepackage[english]{babel}
\usepackage{hyphenat}
\usepackage{amsmath}
\usepackage{amssymb}
\usepackage{mathtools}
\usepackage{mathrsfs}
\usepackage{bbm} 
\usepackage{bm}
\usepackage{slashed}
\usepackage{epstopdf}
\usepackage{xcolor}
\usepackage{booktabs}
\usepackage{float}
\definecolor{lcolor}{rgb}{0.,0.0,0.}
\definecolor{citcolor}{rgb}{0,0.,0.5}
\usepackage[breaklinks,colorlinks,urlcolor=blue,citecolor=blue,linkcolor=blue]{hyperref}
\usepackage{multirow}
\usepackage{ltablex}
\usepackage{soul}

\newcommand{\Hcal}{\mathcal{H}}

\newcommand{\vect}[1]{\boldsymbol{#1}_{\perp}}
\newcommand{\kt}{\boldsymbol{k}}
\newcommand{\kti}{\boldsymbol{k}_i}
\newcommand{\kthi}{\boldsymbol{k}_{hi}}

\newcommand{\Pt}{\boldsymbol{P}}

\newcommand{\lt}{\boldsymbol{l}}

\newcommand{\bt}{\boldsymbol{b}}

\newcommand{\Kt}{\boldsymbol{K}}

\newcommand{\rt}{\vect{r}}

\newcommand{\kht}{\boldsymbol{k_{h\perp}}}
\newcommand{\khtone}{\boldsymbol{k_{h1}}}
\newcommand{\khttwo}{\boldsymbol{k_{h2}}}

\newcommand{\ktone}{\boldsymbol{k_{1}}}
\newcommand{\kttwo}{\boldsymbol{k_{2}}}
\newcommand{\qtone}{\boldsymbol{q_{1}}}
\newcommand{\qttwo}{\boldsymbol{q_{2}}}

\newcommand{\der}{\mathrm{d}}

\newcommand{\Tr}{\mathrm{Tr}}

\allowdisplaybreaks

\begin{document}

\title{Probing gluon saturation with forward heavy-flavor correlations in proton-nucleus collisions}

\author{Farid Salazar}
\email[]{farid.salazar@temple.edu}
\affiliation{Department of Physics, Temple University, Philadelphia, Pennsylvania 19122, USA}
\affiliation{RIKEN-BNL Research Center, Brookhaven National Laboratory, Upton, New York 11973, USA}
\affiliation{Physics Division, Brookhaven National Laboratory, Upton, New York 11973, USA}

\author{Bj\"orn~Schenke}
\email[]{bschenke@bnl.gov}
\affiliation{Physics Division, Brookhaven National Laboratory, Upton, New York 11973, USA}

\author{Wenbin Zhao}
\email[]{WenbinZhao@ccnu.edu.cn}
\affiliation{Institute of Particle Physics and Key Laboratory of Quark and Lepton Physics (MOE), Central China Normal University, Wuhan, 430079, Hubei, China}

\begin{abstract}

We investigate heavy-flavor azimuthal correlations as probes of gluon saturation in high-energy $p$+$p$ and $p$+Pb collisions within the Color Glass Condensate effective theory. Heavy flavor has several advantages: Heavy-quark production is especially sensitive to the small-$x$ gluon distribution of the target, while the quark mass suppresses collinear radiation. Further, the heavy meson carries a large fraction of the momentum of its parent quark, so that the meson pair follows the partonic kinematics more closely than for light hadrons. We present the first calculation of heavy-meson pair correlations over the {\it{full}} azimuthal range that combines small-$x$ evolution from an initial condition constrained by HERA data with Sudakov resummation near the back-to-back limit. It provides a good description of the $D$-$\bar D$ and $b$-hadron pair correlations measured by LHCb. In contrast to the light-flavor case, we predict, in addition to a suppression of the away-side yield in $p$+Pb relative to $p$+$p$ collisions by about $15\%$, a broadening of the peak by approximately $10\%$ to $25\%$ for Gaussian fits near $\Delta\Phi=\pi$, which makes heavy-flavor correlations a promising observable in the search for gluon saturation.

\end{abstract}

\maketitle

\textbf{\textit{Introduction.}} \label{sec:intro}   Establishing the presence of gluon saturation and determining its properties remain central goals in high-energy Quantum Chromodynamics (QCD) research, with implications for hadron structure and high-energy nuclear collisions. At sufficiently high energies, the population of small-$x$ gluons becomes so large that recombination and color screening counteract their production through parton branching, slowing the growth of the gluon density~\cite{Gribov:1983ivg,Mueller:1985wy}. The Color Glass Condensate (CGC) effective theory provides a first-principles description of this high-density regime~\cite{McLerran:1993ni,McLerran:1993ka,McLerran:1994vd,Ayala:1995kg,Ayala:1995hx} (for reviews, see~\cite{Iancu:2003xm,Gelis:2010nm,Kovchegov:2012mbw,Albacete:2014fwa,Morreale:2021pnn}), in which highly occupied small-$x$ gluons are described by classical fields, and the interaction of energetic partons from the projectile with the target by light-like Wilson lines. Large logarithms of $1/x$ are resummed through the evolution of Wilson-line correlators, governed by the Balitsky--JIMWLK hierarchy~\cite{JalilianMarian:1996xn,JalilianMarian:1997dw,Kovner:2000pt,Iancu:2000hn,Iancu:2001ad,Ferreiro:2001qy,Weigert:2005us,Balitsky:1995ub,Kovchegov:1999yj}, which closes for the dipole correlator in the large-$N_c$ limit, yielding the Balitsky--Kovchegov (BK) equation. The CGC has enabled phenomenological studies at HERA, RHIC, and the LHC and will play a central role at the future Electron--Ion Collider~\cite{Morreale:2021pnn}.

Azimuthal angular correlations of hadron pairs at forward rapidity carry information on the saturation dynamics of the dense target \cite{Kharzeev:2004bw,Marquet:2007vb,Albacete:2010pg}. At leading order in a dilute system, the two particles produced in a hard ($2\to2$) partonic scattering have balanced transverse momenta, leading to a pronounced away-side peak near $\Delta\Phi=\pi$. Initial- and final-state radiation generates a momentum imbalance that broadens this peak already in $p$+$p$ collisions, and is resummed in the Sudakov factor~\cite{Mueller:2013wwa,Sun:2014gfa,Marquet:2025jdr}. In the saturation region, multiple scattering imparts additional transverse momentum to the produced partons, leading to further broadening and suppression of the back-to-back correlation. Such modifications have been studied extensively for dihadrons~\cite{Kharzeev:2004bw,Marquet:2007vb,Albacete:2010pg,Lappi:2012nh,Zheng:2014vka,Albacete:2018ruq,Stasto:2018rci,Caucal:2025zkl,Duan:2026jrk}, dijets~\cite{Kotko:2015ura,vanHameren:2019ysa,Al-Mashad:2022zbq}, and photon-hadron and photon-jet pairs~\cite{Jalilian-Marian:2012wwi,Benic:2022ixp}. Energy correlators have been proposed as complementary probes~\cite{Liu:2023aqb,Kang:2023oqj,Kang:2025vjk,Ganguli:2025aqa}.

\begin{figure}
    \centering    
\includegraphics[width=0.9\columnwidth]{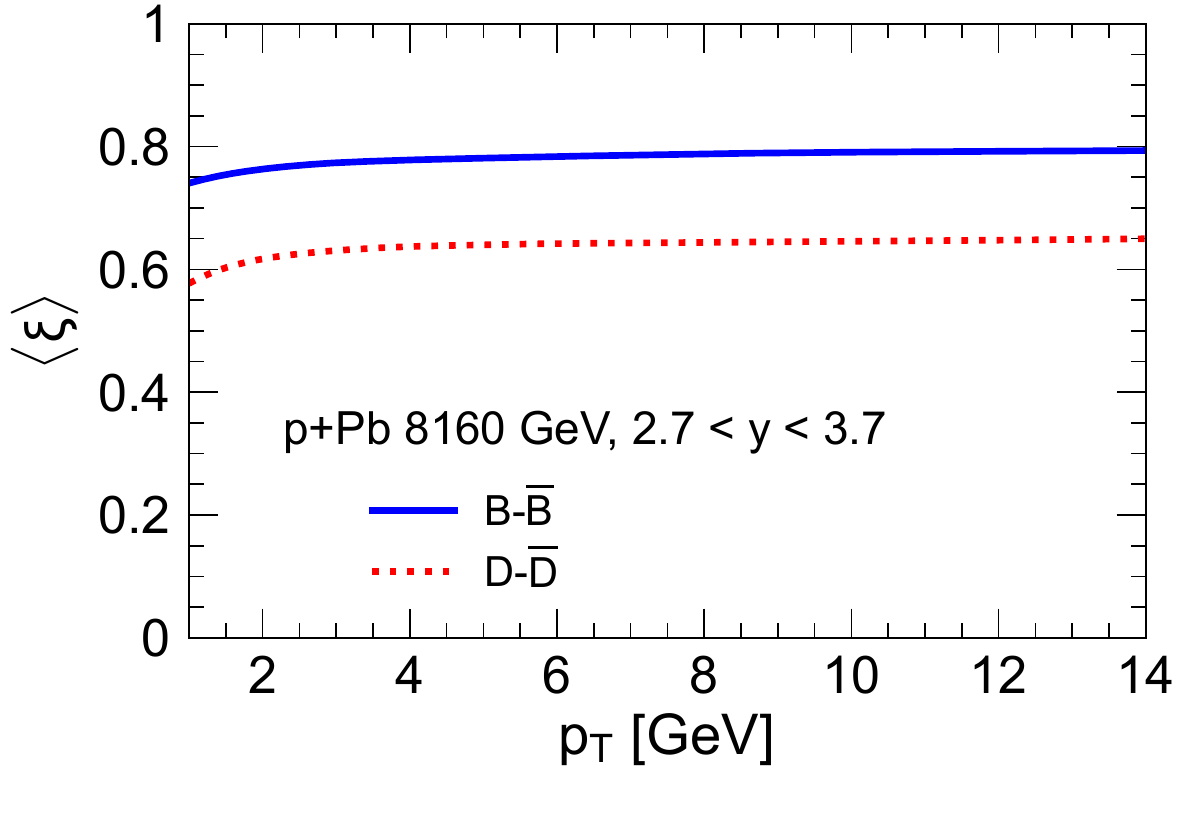}
    \caption{The average momentum fraction $\langle\xi\rangle$ of $D$ and $B$ mesons for $p_T=p_{T1}=p_{T2}$ in $p$+Pb collisions at
$\sqrt{s_{\mathrm{NN}}}=8.16$~TeV in the rapidity bin $2.7<y<3.7$.}
    \label{fig:z1}
\end{figure}

Motivated by our study of light-hadron correlations~\cite{Caucal:2025zkl}, we investigate correlations of heavy mesons, whose production is dominated by gluon-initiated processes and therefore sensitive to the small-$x$ gluon distribution of the target. Near $\Delta\Phi\simeq\pi$ the decorrelation receives contributions from Sudakov radiation, hadronization and transverse-momentum broadening in the dense target. In our light-hadron study we found that the away-side width is controlled by Sudakov radiation and hadronization, so that the width ratio between $p$+A and $p$+$p$ collisions remains close to unity at RHIC and LHC energies, while the yield is still suppressed~\cite{Caucal:2025zkl}, in agreement with the STAR data~\cite{STAR:2021fgw}. For heavy quarks both effects are reduced: the dead cone reduces the final-state collinear radiation~\cite{delCastillo:2020omr,Dai:2026hso,Marquet:2025jdr,Gao:2026azd,Shao:2026doo}, and the heavy meson carries a large fraction of the momentum of its parent quark~\cite{Peterson:1982ak,Cacciari:2005uk} (see Fig.~\ref{fig:z1}), so that the meson pair follows the partonic kinematics more closely than for light hadrons. This makes charm and bottom correlations better suited for disentangling saturation effects from radiation and hadronization. Alternatively, the Sudakov contribution can be reduced through the choice of observable~\cite{Ke:2026ytl,Lin:2026typ}.

In this work, we present the first calculation of heavy-flavor pair correlations over the full azimuthal range that combines small-$x$ evolution from an initial condition constrained by HERA data~\cite{Albacete:2010sy} with Sudakov resummation near the back-to-back limit. Earlier small-$x$ calculations covered the full azimuthal range without Sudakov resummation, reproducing the LHCb $D$-$\bar D$ correlation only with an initial condition disfavored by HERA data~\cite{Fujii:2013yja}, while more recent studies include Sudakov resummation but are restricted to the back-to-back region~\cite{Marquet:2025jdr,Gao:2026azd}. We extract the double and single logarithms of the heavy-quark pair, together with their dead-cone modification, from the factorization for massive quarks formulated in soft-collinear effective theory (SCET)~\cite{delCastillo:2020omr,vonKuk:2023jfd}, and find that the double logarithms agree with those of an earlier heavy-flavor study~\cite{Marquet:2025jdr} (Supplemental Material~\ref{app:sudakov}). We turn off the perturbative Sudakov factor smoothly toward the near side, with a matching between the resummed and unresummed leading-order CGC results similar to the one used for heavy-quark spectra~\cite{Cacciari:1998it,Cacciari:2001td}, while the nonperturbative Sudakov factor, which models the transverse-momentum broadening from hadronization, is kept over the full angular range. Fragmentation into heavy mesons is treated with exact kinematics for massive quarks and hadrons. 

LHCb has measured prompt $D$- and $B$-meson production in $p$+$p$ and $p$+Pb collisions~\cite{LHCb:2016ikn,LHCb:2017yua,LHCb:2013vjr,LHCb:2019avm}, as well as open-charm pair production in $p$+$p$ and $p$+Pb collisions and correlations of $b$-hadron pairs, reconstructed through their decays into $J/\psi$ mesons, in $p$+$p$ collisions~\cite{LHCb:2012aiv,LHCb:2017bvf,LHCb:2020jse}. We confront our calculation with these data, using the same parameters in $p$+$p$ and $p$+Pb collisions, and present predictions for $D$-$\bar D$ and $B$-$\bar B$ correlations in $p$+Pb collisions, focusing on the nuclear modification of the yield and width of the away-side peak. For heavy mesons with $p_T\sim2$-$8$~GeV and $y\sim 2.5$-$4.5$ the cross section is dominated by $x_A\sim10^{-5}$-$10^{-4}$, where the saturation scale of the lead nucleus is expected to be comparable to the transverse momenta of the produced heavy quarks.

\textbf{\textit{Theoretical formalism.}} \label{sec:theory-sec} In the dilute-dense approximation and in the large $N_c$ limit, the leading order differential cross-section for heavy-quark pair production in proton-nucleus collisions within the CGC effective theory is given by \cite{Blaizot:2004wv,Fujii:2006ab,Fujii:2013yja,Kovchegov:2006qn,Altinoluk:2015vax}
\begin{widetext}
    \begin{align}
    & \frac{\der \sigma^{pA \to Q\bar{Q}X}}{\der^2 \ktone \der^2 \kttwo \der y_1 \der y_2}  = x_p g(x_p,\mu^2) S_\perp \int \frac{\der^2 \lt}{(2\pi)^2}   \Hcal(\Pt,\Pt+z\Kt-\lt,z) (z\Kt-\lt)^2 F(\lt;x_A) F(\Kt-\lt;x_A)\,, \label{eq:double-inclusive-CGC-large-Nc}
\end{align}
\end{widetext}
where $S_\perp$ is the effective transverse area of the target, $\mu^2$ the factorization scale, and $\ktone, y_1$ ($\kttwo, y_2$) are the transverse momentum and rapidity of the heavy quark (antiquark), respectively. The partonic momentum imbalance is $\Kt = \ktone + \kttwo$, and the relative momentum $\Pt = \bar{z} \ktone - z \kttwo$, with $z= m_{T1} e^{y_1}/(m_{T1} e^{y_1} + m_{T2} e^{y_2})$ and $\bar{z}=1-z$, where $m_{Ti} = \sqrt{\kti^2 + m^2}$ with $m$ the heavy quark mass. The longitudinal momentum fractions probed in the proton and nucleus are respectively
\begin{align}
    x_p=(m_{T1} e^{y_1} + m_{T2} e^{y_2})/\sqrt{s} \,, \\
    x_A=( m_{T1} e^{-y_1} + m_{T2} e^{-y_2})/\sqrt{s} \,.
\end{align}
At forward rapidities $x_A \ll 1$, while $x_p \gg x_A$, justifying the dilute-dense approximation~\cite{Dumitru:2005gt,Altinoluk:2011qy} (see~\cite{Fujii:2026ccu} for a comparison with the dense-dense framework).

We introduced the Fourier transform of the dipole correlator
\begin{align}
    F(\lt;x) = \int \frac{\der^2 \rt}{(2\pi)^2} e^{-i\lt \cdot \rt} \frac{1}{N_c}\left \langle \Tr\left[V(\rt) V^\dagger(\boldsymbol{0})\right] \right  \rangle_x \,,
\end{align}
and the perturbative factor $\Hcal(\qtone,\qttwo,z)  = \left[ H(\qtone,\qtone,z) - 2 H(\qtone,\qttwo,z) + H(\qttwo,\qttwo,z) \right]/(\qtone-\qttwo)^2$ with 
\begin{align}
    &H(\qtone, \qttwo,z) = \alpha_s  z\bar{z} \left[\frac{ (z^2 + \bar{z}^2) (\qtone \cdot \qttwo) + m^2}{(\qtone^2 + m^2) (\qttwo^2 + m^2)} \right]\,.
\end{align}

\begin{widetext}
The dihadron production cross section including the Sudakov factor can be written as 
\begin{align}
    \frac{\der \sigma^{pA \to H_1 H_2 X}}{\der^2 \khtone \der^2 \khttwo \der y_{h1} \der y_{h2}}  & = S_\perp \int^1_{\xi_{1,\mathrm{min}}} \!\!\!\!\!\! \der \xi_1 \mathcal{J}_1 \int^1_{\xi_{2,\mathrm{min}}} \!\!\!\!\!\! \der \xi_2 \mathcal{J}_2 \int \frac{\der^2 \lt}{(2\pi)^2} \Hcal(\Pt,\Pt+z\Kt-\lt,z)  \int \frac{\der^2 \Kt'}{(2\pi)^2} W(\Kt'-\Kt, \xi_1,\xi_2,x_p, \mu^2) \nonumber \\ & \times (z\Kt'-\lt)^2 F(\lt;x_A) F(\Kt'-\lt;x_A) \,.
    \label{eq:dihadron-sudakov}
\end{align}
\end{widetext}
Here $\Kt'$ is the transverse momentum transferred by the target, which the Sudakov radiation shifts to the measured imbalance $\Kt$. As in transverse-momentum-dependent (TMD) factorization~\cite{Dominguez:2011wm,Mueller:2013wwa,Caucal:2026yzl}, the convolution with $W$ acts only on the target distributions, while the hard factor is evaluated at the measured kinematics, so that for $|\Kt|,|\Kt'|,Q_s\ll|\Pt|$ Eq.~\eqref{eq:dihadron-sudakov} reduces to the TMD result~\cite{Marquet:2025jdr,Gao:2026azd}. Away from this limit, the hard factor retains the kinematic-twist corrections of the leading-order CGC result~\cite{Kotko:2015ura,Altinoluk:2019fui,Altinoluk:2021ygv}. Each heavy meson is taken on its mass shell, collinear to its parent quark, and carrying a fraction $\xi_i$ of the quark light-cone momentum along the quark direction, so that the quark momenta in Eq.~\eqref{eq:dihadron-sudakov} are fixed by the meson momenta and $\xi_i$, with the Jacobian $\mathcal{J}_i = |\partial(\kti,y_i) /\partial (\kthi,y_{hi})|$ given in Supplemental Material~\ref{app:kinematics}. The integrations over $\xi_i$ are restricted to $x_p,x_A<1$.

Following~\cite{Benic:2022ixp,Caucal:2025zkl}, we define
\begin{align}
    & W(\lt, \xi_1,\xi_2,x,
\mu^2)  
    = \int \der^2 \bt\ e^{i \lt \cdot \bt} e^{-S_{\mathrm{P}}(\bt,\mu^2)} e^{-S_{\mathrm{NP}}(\bt,\mu^2)}  \nonumber \\
    & \times xg(x,\mu_b^2)  D_{H_1/Q}(\xi_1,\mu_b^2) D_{H_2 /\bar{Q}}(\xi_2,\mu_b^2) \,,
\end{align}
with the perturbative Sudakov term
\begin{align}
\label{eq:sudakov}
    & S_{\mathrm P}(\bt,\mu^2) =  \int_{\mu_b^2}^{\mu^2} \frac{\der \bar{\mu}^2}{\bar{\mu}^2} \frac{\alpha_s(\bar{\mu})}{2\pi} \Bigg\{  2C_A  \ln\left( \frac{\mu^2}{\bar{\mu}^2} \right) -2\beta_0  \nonumber  \\
    &  + 2 C_F \left[ \ln\left(\frac{\mu^2}{\bar{\mu}^2}\right) - \theta(m^2-\bar{\mu}^2) \ln\left(\frac{m^2}{\bar{\mu}^2}\right)\right] \nonumber \\
    & - 2C_F \left[ \frac{3}{2} - \frac{1}{2} \theta(m^2-\bar{\mu}^2)  \right] \Bigg\}\,,
\end{align}
and $\beta_0=(11N_c-2N_f)/12$, with $N_f=3$, 4 and 5 active flavors below $m_c$, between $m_c$ and $m_b$, and above
$m_b$, respectively. The hard scale is $\mu^2=x_px_As=M^2+\Kt^2$, where
$M^2=(\Pt^2+m^2)/(z\bar z)$ is the invariant mass of the heavy-quark pair. The step functions implement the dead-cone effect: below the quark mass collinear radiation is suppressed, which freezes the double logarithm at $m$ and modifies the single logarithm. The double logarithms agree with those of earlier heavy-flavor studies~\cite{Marquet:2025jdr,Gao:2026azd}, and the single logarithms of the heavy-quark pair, extracted in Supplemental Material~\ref{app:sudakov} from the factorization for massive quarks~\cite{delCastillo:2020omr,vonKuk:2023jfd}, agree with those of the TMD calculation for massive quarks~\cite{Gao:2026azd} up to a color-correlated soft term suppressed by $1/N_c^2$, which we neglect in the large-$N_c$ limit. 

\begin{figure}
    \centering    
\includegraphics[width=0.95\columnwidth]{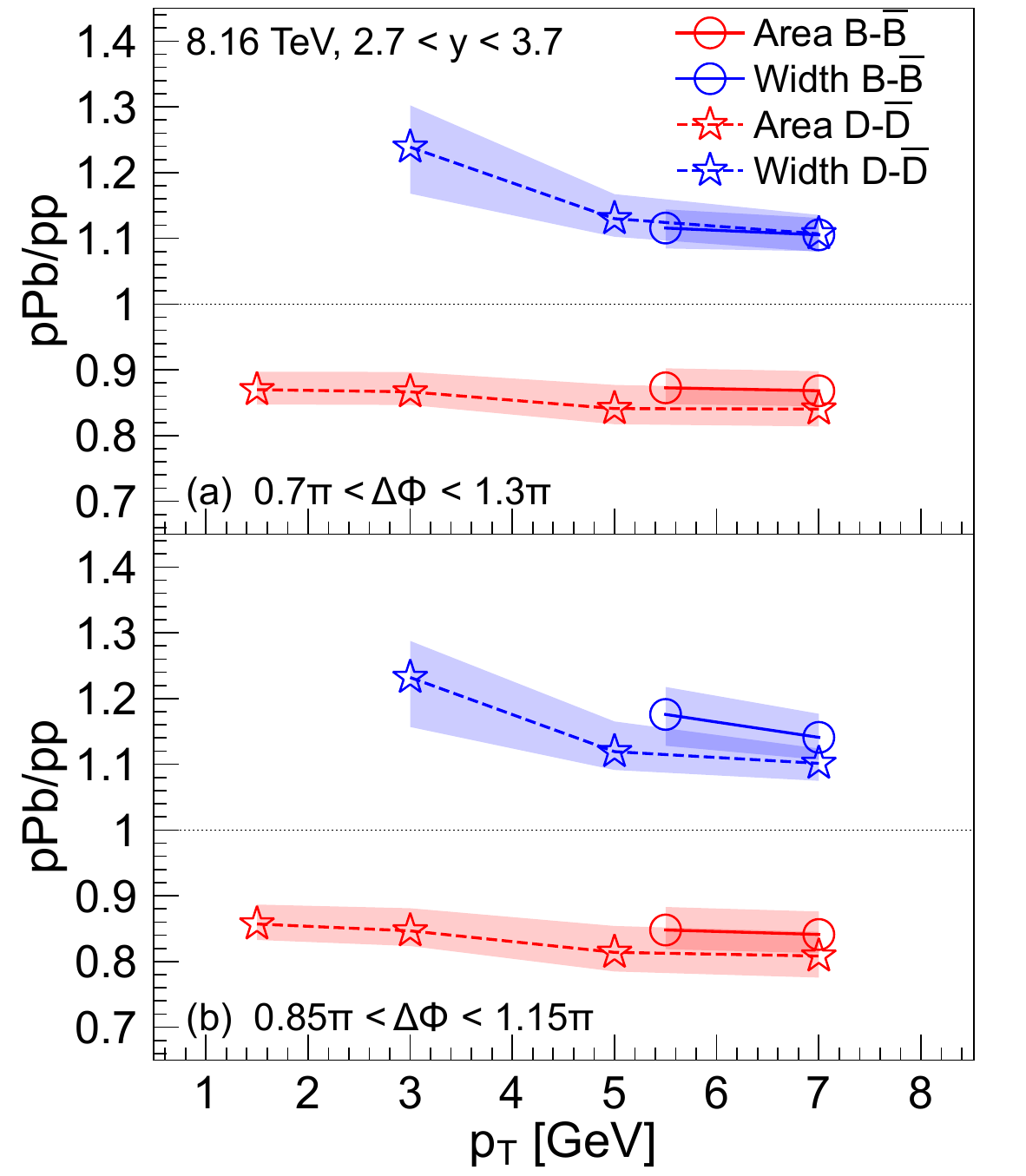}
    \caption{Ratios of the width and area of the away-side peak in $p$+Pb to those in $p$+$p$ collisions, computed at $\sqrt{s_{\mathrm{NN}}} = 8.16$ TeV with both mesons in the rapidity bin $2.7 < y < 3.7$ and in the same $p_T$ bin, for $D$-$\bar D$ and $B$-$\bar B$ pairs, as a function of $p_T$. The area is the integral of $CP(\Delta\Phi)$ over $0.7\pi<\Delta\Phi<1.3\pi$ (a) or $0.85\pi<\Delta\Phi<1.15\pi$ (b), and the width that of a Gaussian fit on the same window, not shown for $1<p_T<2$~GeV. The bands are obtained by varying $Q_{s0,{\rm Pb}}^2$ by factors of 0.8, 1.0, and 1.2.}
    \label{fig:ratio}
\end{figure}

To avoid the infrared region at large $b$ we use the Collins-Soper-Sterman (CSS) prescription~\cite{Collins:1984kg}, $\mu_b=c_0/b_*$ with $c_0=2e^{-\gamma_E}$ and $b_*=b/\sqrt{1+b^2/b_{\max}^2}$, so that $\mu_b>c_0/b_{\max}$, and we set $\mu_b=\mu$ when $\mu_b>\mu$, so that $S_{\rm P}=0$ at short distances. We parametrize the remaining long-distance contribution~\cite{Sun:2014dqm,Sun:2014gfa,Chen:2016vem} by
\begin{align}
     S_{\mathrm{NP}}(\bt,\mu^2)  = C_{\rm NP}\left(\frac{C_A}{C_F} + 2   \right) \times \notag \\ & \hspace{-1.8cm}\left[ \frac{g_1}{2} \bt^2 + \frac{g_2}{8} \ln\left( \frac{\mu^2}{Q_0^2} \right) \ln\left( \frac{\bt^2}{b_*^2}\right) \right]\,,
\end{align}
where the factor $(C_A/C_F+2)$ follows from Casimir scaling for the incoming gluon and the two outgoing heavy quarks. We use the values of our light-hadron study~\cite{Caucal:2025zkl}, $g_1=0.212$~GeV$^2$, $g_2=0.84$, $Q_0^2=2.4$~GeV$^2$, $b_{\max}=1.5$~GeV$^{-1}$, and $C_{\rm NP}=2.3$, fitted there to the away-side width of forward di-$\pi^0$ correlations in $p$+$p$ collisions, assuming that the nonperturbative input is the same for light and heavy flavors. The ratios between $p$+Pb and $p$+$p$ collisions of Fig.~\ref{fig:ratio} change by less than $2\%$ between $C_{\rm NP}=1$ and $2.3$.

\begin{figure*}
    \centering    
\includegraphics[width=1\linewidth]{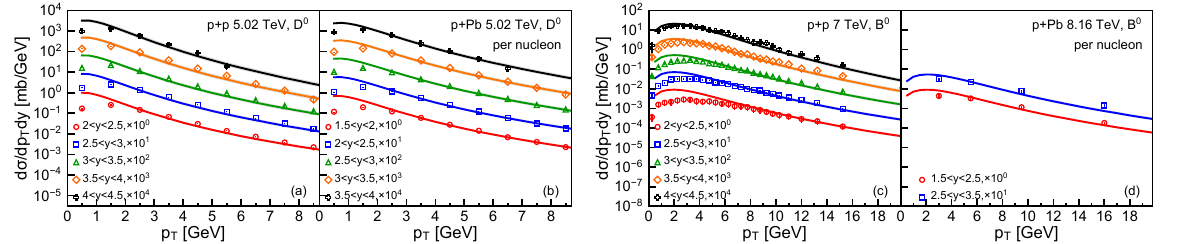}
    \caption{The spectra of $D^0$ mesons (left two panels) and $B^0$  (right two panels)  in $p$+$p$ and $p$+Pb collisions at $\sqrt{s_{\mathrm{NN}}} = 5.02$ TeV ($D^0$), and in $p$+$p$ collisions at 7 TeV and $p$+Pb collisions at 8.16 TeV ($B^0$), in different rapidity bins.  The $p$+Pb spectra are divided by $A = 208$ to give per-nucleon cross sections. The bands show the uncertainty from varying the factorization scale by factors of 0.5 and 2. The curves include the factors $K_D=4.8$ and $K_B=3.1$, fitted to the data above $p_T=2$~GeV. LHCb data from~\cite{LHCb:2016ikn,LHCb:2017yua,LHCb:2013vjr,LHCb:2019avm}.}
    \label{fig:spectra}
\end{figure*}

\begin{figure*}
    \centering    
\includegraphics[width=2.1\columnwidth]{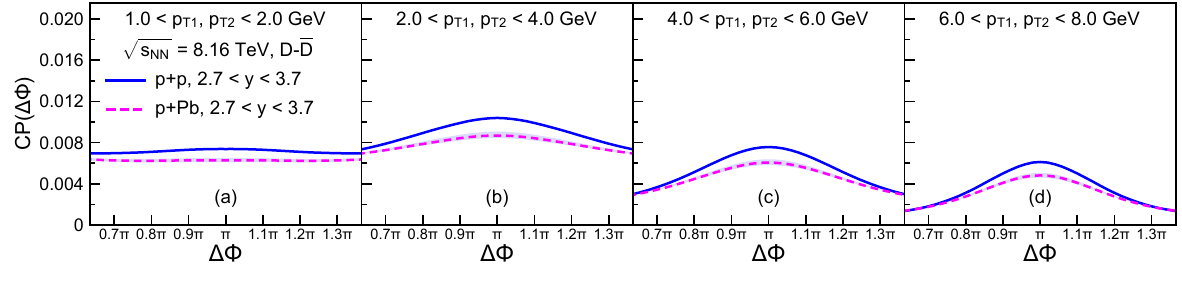}
    \caption{The coincidence probability $CP(\Delta\Phi)$ of $D$-$\bar D$ pairs in $p$+$p$ and $p$+Pb collisions at $\sqrt{s_{\mathrm{NN}}}=8.16$~TeV, with both mesons in the rapidity bin $2.7<y<3.7$ and in the same $p_T$ bin, for $1<p_T<2$, $2<p_T<4$, $4<p_T<6$ and $6<p_T<8$~GeV. The bands for $p$+Pb collisions are obtained by varying $Q_{s0,{\rm Pb}}^2$ by factors of 0.8, 1.0, and 1.2.}
    \label{fig:DD_prediction}
\end{figure*}

\begin{figure}
    \centering    
\includegraphics[width=0.9\columnwidth]{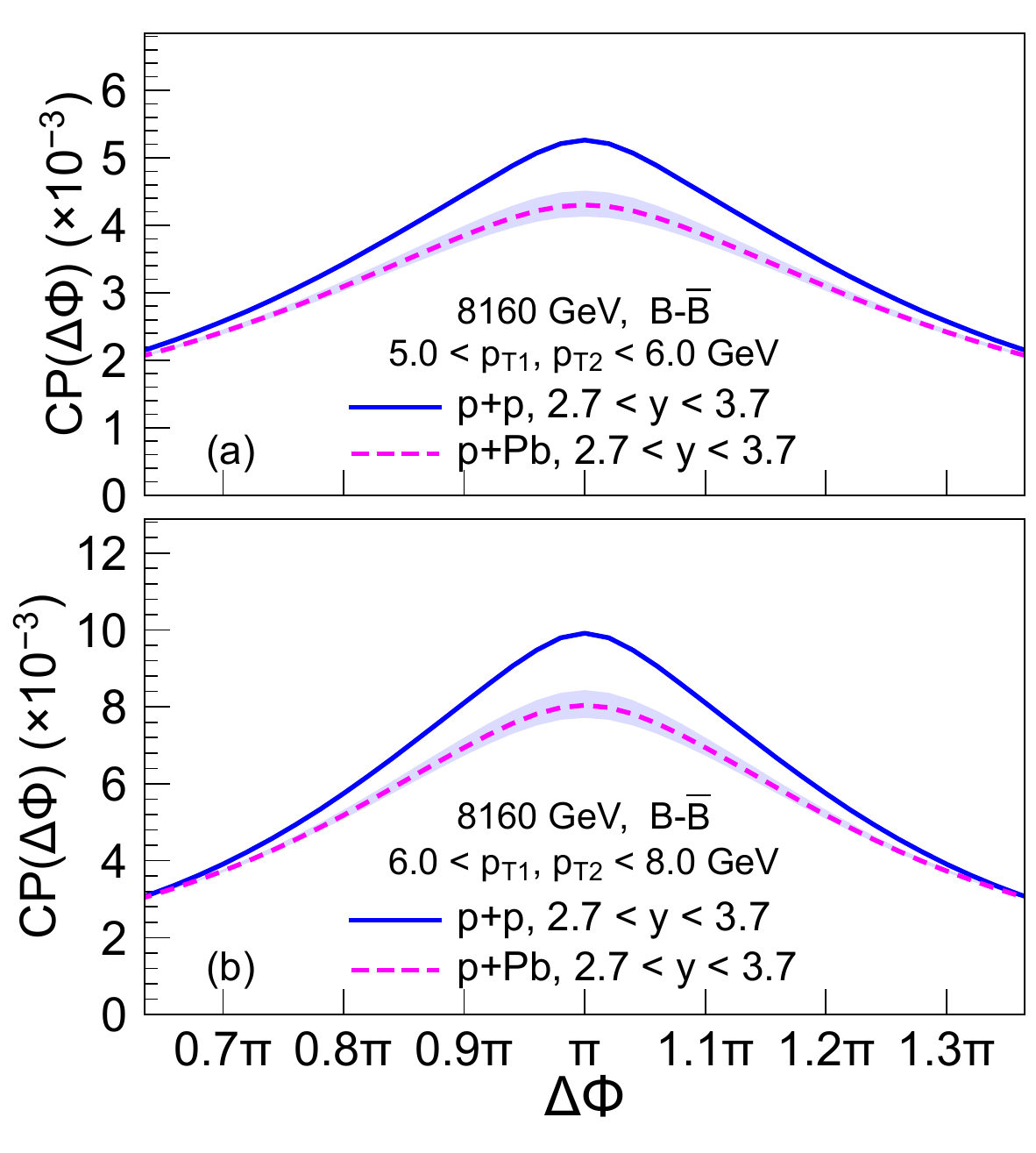}
    \caption{The coincidence probability $CP(\Delta\Phi)$ of $B$-$\bar B$ meson pairs in $p$+$p$ and $p$+Pb collisions at $\sqrt{s_{\mathrm{NN}}}=8.16$~TeV, with both mesons in the rapidity bin $2.7<y<3.7$ and in the same $p_T$ bin, for $5<p_T<6$ and $6<p_T<8$~GeV. The bands are obtained as in Fig.~\ref{fig:DD_prediction}.}
    \label{fig:BB_prediction}
\end{figure}

Since the perturbative Sudakov factor applies only near the back-to-back limit, we replace $W$ in Eq.~\eqref{eq:dihadron-sudakov} by
\begin{align}
    W \to G(R)\,W + \left[1-G(R)\right] W_{\rm NP}\,,
\end{align}
where $W_{\rm NP}$ is $W$ with $S_{\rm P}=0$ and with the gluon distribution and fragmentation functions at $\mu^2$. 
\begin{equation}
    G(R) = \frac{R}{R + c^{2}}\,, ~~ {\rm with~} R = \frac{M^2}{\Kt'^2}\,,
\label{eq:matching}
\end{equation}
is evaluated for each configuration in the integrand, with its own $M$ and $\Kt'$, similar to the matching used for heavy-quark spectra~\cite{Cacciari:2001td}. This prescription gives the resummed result for $|\Kt'|\ll M$, where the logarithms of $M^2/\Kt'^2$ are large, and the leading-order CGC result smeared by the nonperturbative factor for $|\Kt'|\gg M$, where this factor remains important~\cite{Barata:2024wsu,Guo:2025zwb,Kang:2025zto,Guo:2026bhd}. We use $M^2$ rather than $\mu^2=M^2+\Kt^2$ in $R$, since $\mu^2/\Kt^2$ never falls below one, and vary $c=0.5$, $1.0$ and $2.0$ to estimate the matching uncertainty. We use $c=1.0$ in Figs.~\ref{fig:ratio} to \ref{fig:BB_prediction}.

We adopt the CT18NNLO PDF parametrization~\cite{Hou:2019efy} for the collinear gluon distribution and the Kneesch-Kniehl-Kramer-Schienbein (KKKS08) fragmentation function (FF)~\cite{Kneesch:2007ey} for $c\to D$. For $B$-$\bar B$ pairs reconstructed through $b\to J/\psi$ decays we construct, as in~\cite{Bolzoni:2013tca}, an effective FF
\begin{equation}
D_{b\to J/\psi}(\xi,\mu)=\int_{\xi}^{1}\frac{\mathrm{d}u}{u}\,
D_{B/b}\!\left(\frac{\xi}{u},\mu\right)
\frac{1}{\Gamma_B}\frac{\mathrm{d}\Gamma}{\mathrm{d}u}(u,P_B)\,,
\label{eq:DbJpsi}
\end{equation}
where $D_{B/b}$ is the Kniehl-Kramer-Schienbein-Spiesberger (KKSS) $b\to B$ FF~\cite{Kniehl:2007erq}, and $\mathrm{d}\Gamma/\mathrm{d}u$ is the $B\to J/\psi+X$ decay distribution in the fraction $u$ of the $B$ three-momentum $P_B=\big[p_{J/\psi,T}^2+(p_{J/\psi,T}^2+m_{J/\psi}^2)\sinh^2 y_{J/\psi}\big]^{1/2}/u$ carried by the collinear $J/\psi$, computed in leading-order nonrelativistic QCD including the feed-down from $\chi_{cJ}$ and $\psi(2S)$~\cite{Kniehl:1999vf,Bolzoni:2013tca}, with $\Gamma_B=1/\tau_B$ and $\tau_B=1.61$~ps. The masses of the $c$ and $b$ quarks are $1.28$~GeV and $4.5$~GeV.

Single inclusive production is obtained by integrating over the antiquark and convoluting with the fragmentation function, 
\begin{align}
    & \frac{\der \sigma^{pA \to H X}}{\der^2 \kht \der y_{h} } = 2 S_\perp \int^1_{\xi_{\mathrm{min}}} \!\!\!\!\!\! \der \xi \int \der y_2 \mathcal{J} D_{H/Q}(\xi,\mu^2) x_p g(x_p,\mu^2) \nonumber \\
    & \times \int  \der^2 \lt'  \int \frac{\der^2 \lt}{(2\pi)^2} \ \Hcal\left( \ktone-\lt, \ktone - z (\lt' +\lt),z \right)   \nonumber \\
     & \times (z\lt'-\bar{z}\lt)^2 F(\lt;x_A) F(\lt';x_A) \,, \label{eq:single-inclusive-analytic}
\end{align}
where $\mathcal{J} = |\partial(\kt,y) /\partial (\kht,y_{h})|$ is the Jacobian and the factor of two accounts for the charge-conjugate channel, since the LHCb cross sections include $D^0$ and $\bar D^0$ ($B^0$ and $\bar B^0$). We integrate numerically over the momentum and rapidity of the unobserved antiquark, so that $x_p$, $x_A$ and $\mu^2$ retain their exact kinematics, as in the pair cross section.

\textbf{\textit{Results.}}
\label{sec:Results} We first compare our results with the LHCb single inclusive spectra and self-normalized $D$-$\bar D$ and $B$-$\bar B$ correlations, and then present predictions for the nuclear modification of the away-side peak, through the ratios of its width and area in $p$+Pb and $p$+$p$ collisions. Both are extracted from the coincidence probability, the number of pairs per trigger meson, $CP(\Delta\Phi)=N_{\rm pair}(\Delta\Phi)/N_{\rm trig}$~\cite{STAR:2021fgw,Caucal:2025zkl}, which we show in Supplemental Material~\ref{app:dihadron}. 

The LHCb single inclusive $D^0$ and $B^0$ spectra~\cite{LHCb:2016ikn,LHCb:2017yua,LHCb:2013vjr,LHCb:2019avm} have been previously studied in the CGC~\cite{Fujii:2013yja,Fujii:2017rqa,Ma:2018bax}. For the dipole correlator we use the solution of the running-coupling BK equation with the MV$^\gamma$ initial condition fitted to HERA data~\cite{Albacete:2010sy}, as in our light-hadron study~\cite{Caucal:2025zkl}. We set the effective transverse area of the proton to $S_{\perp,p}=\sigma_0/2=15.13$~mb, with $\sigma_0$ from the AAMQS fit~\cite{Albacete:2010sy}. 
Since we neglect the impact-parameter dependence of the target, $S_{\perp,\mathrm{Pb}}$ is an effective normalization rather than a geometric area. We fix it through $S_{\perp,\mathrm{Pb}}Q^2_{s0,\mathrm{Pb}}=A\,S_{\perp,p}Q^2_{s0,p}$, as suggested by impact-parameter-dependent calculations of minimum-bias cross sections~\cite{Deganutti:2023qct}. For the MV$^\gamma$ initial condition this gives $p$+Pb cross sections close to, but not exactly, $A$ times the $p$+$p$ ones at large transverse momentum, with a nuclear modification factor of $0.96$ to $1.04$ for $D^0$ mesons at $p_T=8$ to $10$~GeV. We use $Q^2_{s0,{\rm Pb}}/Q^2_{s0,p}=2.75$~\cite{Caucal:2025zkl} (see also a recent global fit of the nuclear dipole amplitude~\cite{Dai:2026nzp,Dai:2026rgl}). The normalization factors fitted to the spectra are $K_D=4.8$ for $D^0+\bar{D}^0$ production in $p$+$p$ and $p$+Pb collisions at 5.02~TeV, and $K_B=3.1$ for $B^0+\bar{B}^0$ production in $p$+$p$ collisions at 7~TeV and $p$+Pb collisions at 8.16~TeV. These factors absorb the uncertainty in the effective transverse area and the higher-order corrections, which in collinear factorization are large for heavy-flavor production at low $p_T$~\cite{Nason:1987xz,Nason:1989zy,Cacciari:1998it,Kniehl:2012ti}. They cancel in the self-normalized correlations, in the coincidence probability and in the ratios between $p$+Pb and $p$+$p$ collisions.

With the same parameters, the spectra in Fig.~\ref{fig:spectra} are well described for $p_T\gtrsim2$~GeV in both $p$+$p$ and $p$+Pb collisions. At lower $p_T \lesssim m$, the calculation overshoots the data, as also found in an earlier CGC study~\cite{Ma:2018bax}, and the leading-order result depends strongly on the treatment of the fragmentation functions and of the scales, as in collinear factorization at low $p_T$~\cite{Cacciari:1998it,Cacciari:2012ny,Kniehl:2012ti,Helenius:2018uul,Helenius:2023wkn}.

In contrast to the single inclusive yield, where the momentum transferred by the target is shared with the unobserved antiquark, the imbalance $\Kt$ of a measured pair gives more direct access to the nuclear momentum transfer~\cite{Kharzeev:2004bw,Dominguez:2010xd,Caucal:2025zkl}. Heavy-meson pair correlations have recently been studied in the CGC with Sudakov resummation within transverse-momentum-dependent factorization~\cite{Marquet:2025jdr,Gao:2026azd}, the appropriate framework near the back-to-back limit, where it describes the shape of the measured away-side peak. Since the LHCb correlations are normalized to their integral over the full azimuthal range, these comparisons are performed after normalizing the calculation and the data within a window around $\Delta\Phi=\pi$. Our calculation extends this description beyond the back-to-back limit to the full azimuthal range, so that the near side, the transition region and the away-side peak are compared with the data with a single normalization, which also tests the relative weight of the near-side and away-side peaks. In addition, it uses DGLAP-evolved fragmentation functions with an exact parton-to-hadron map for massive quarks and hadrons, and the coincidence probability is computed with the same kinematics for the trigger and the pair.

\begin{figure*}
    \centering    
\includegraphics[width=0.9\columnwidth]{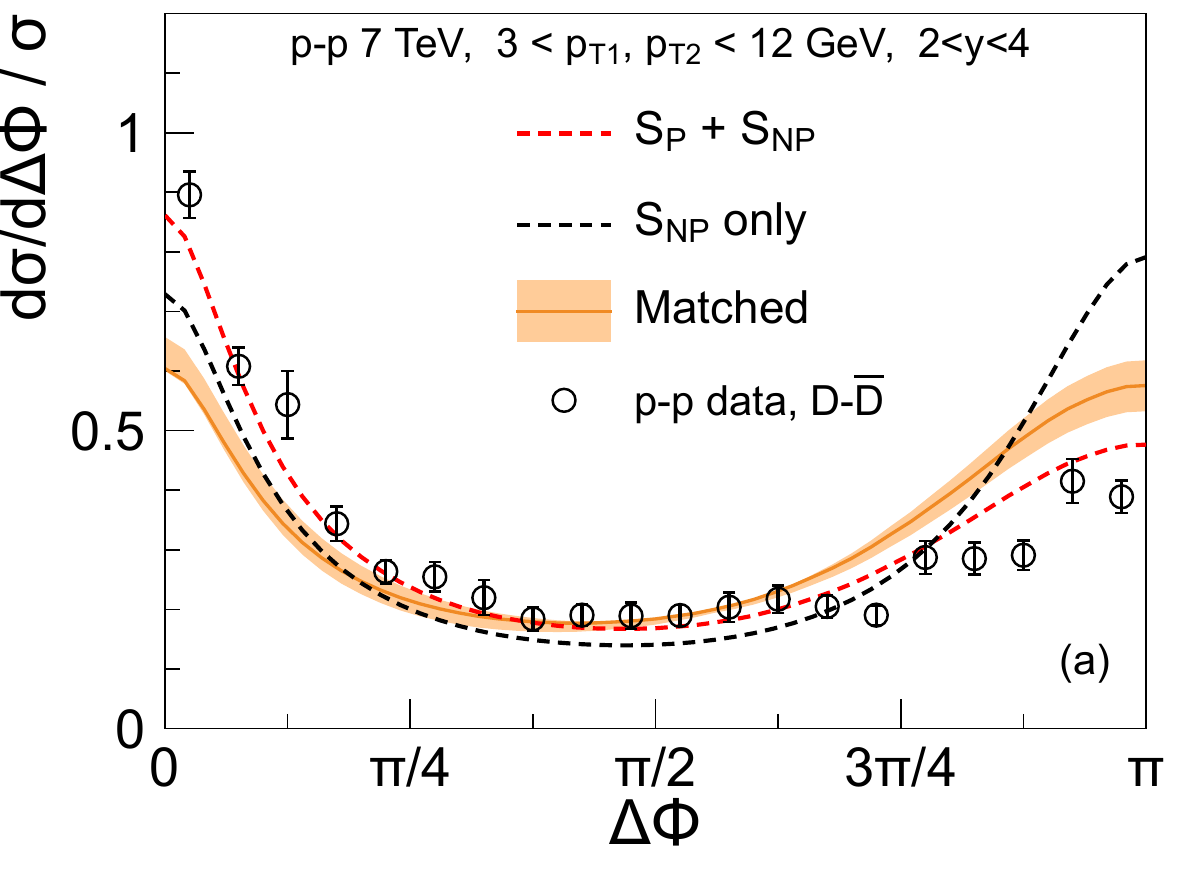}
\includegraphics[width=0.9\columnwidth]{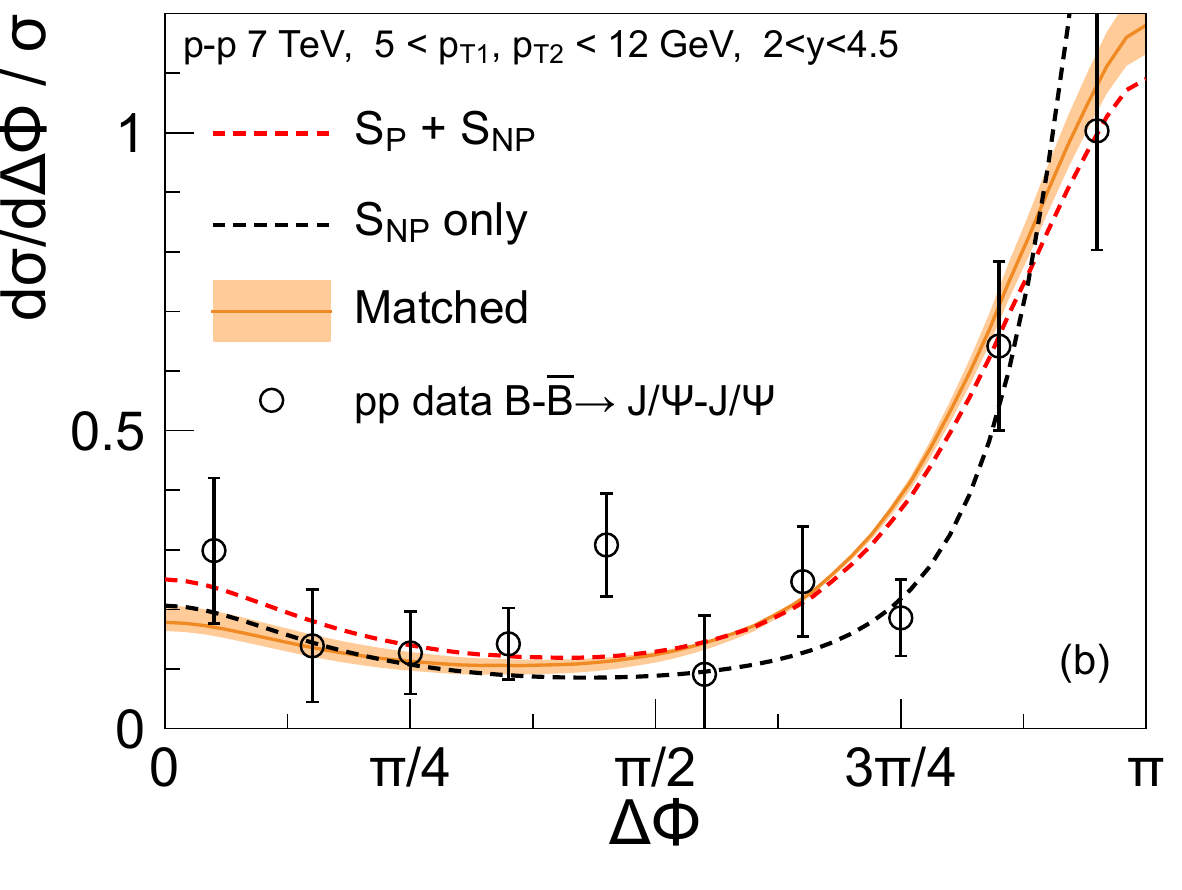}
\includegraphics[width=0.9\columnwidth]{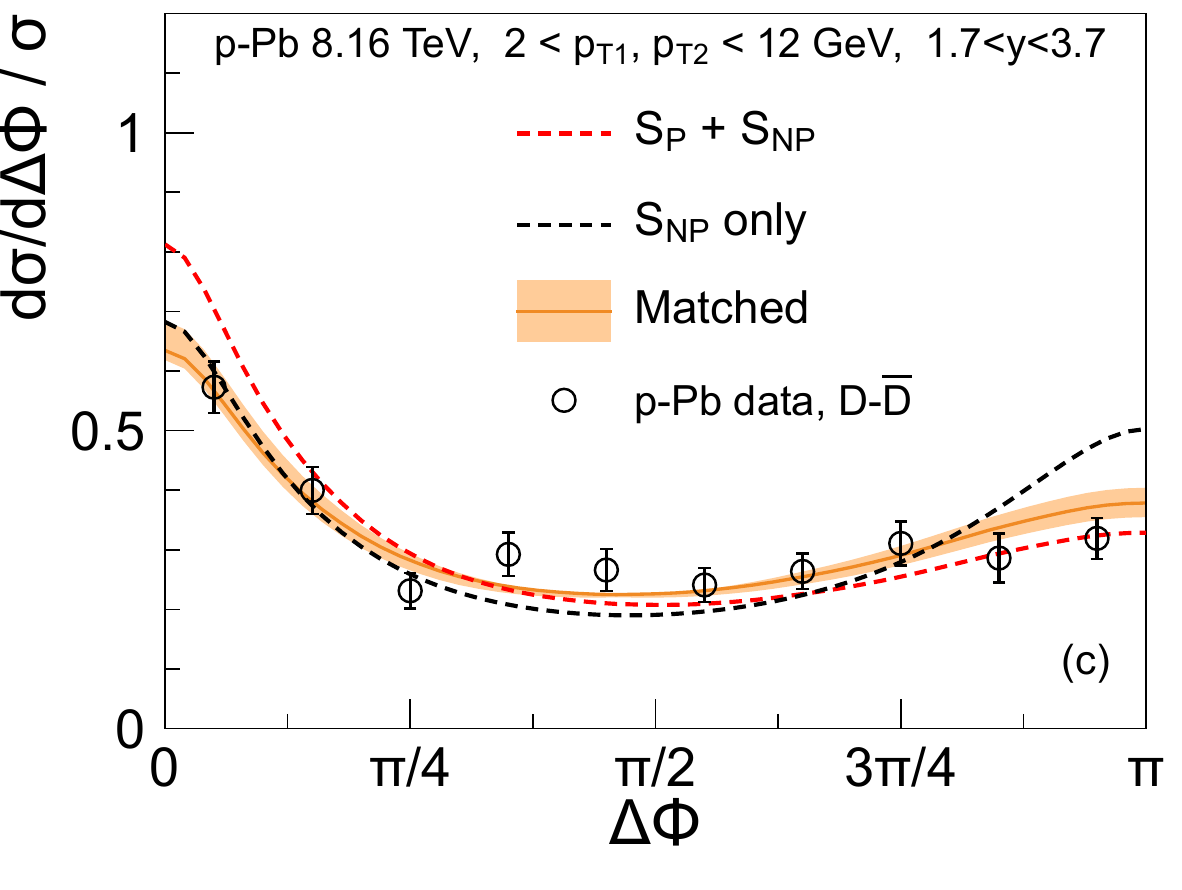}
\includegraphics[width=0.9\columnwidth]{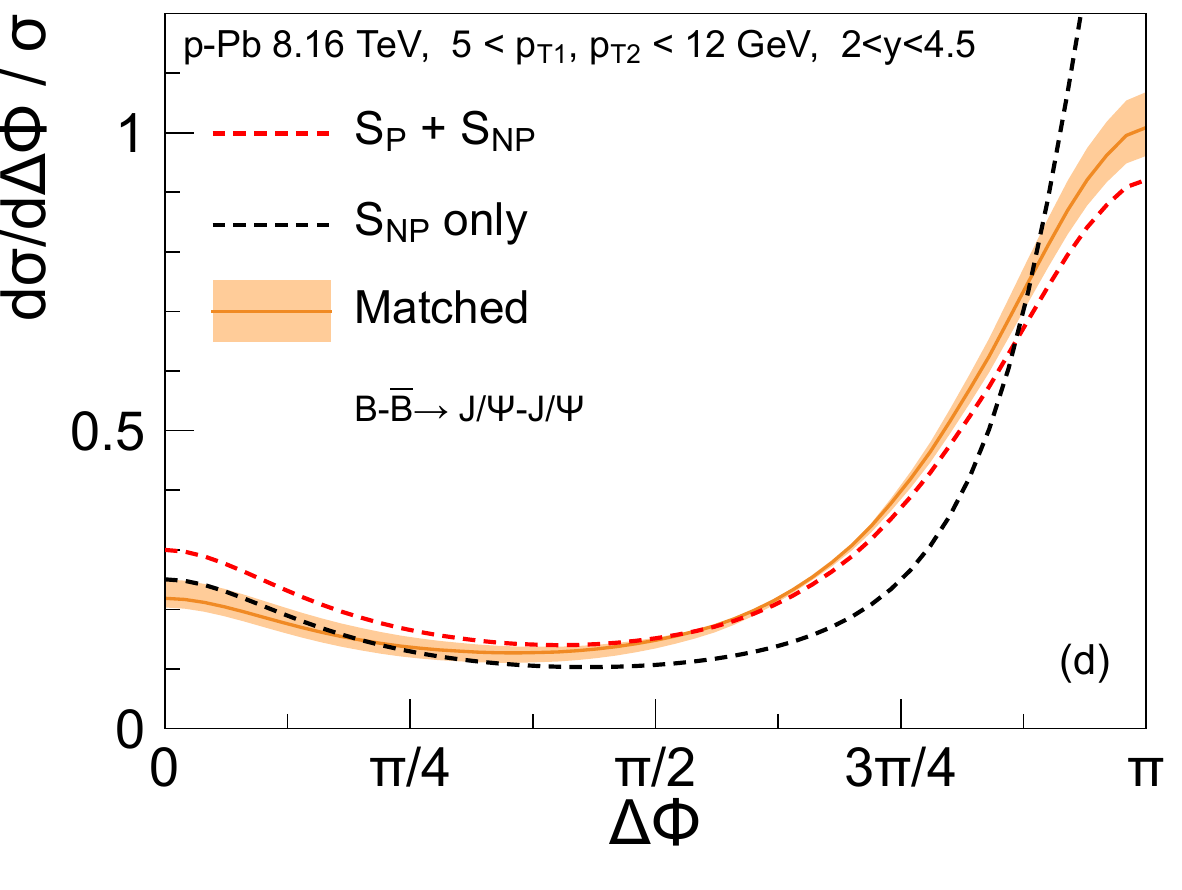}
    \caption{Self-normalized azimuthal correlations over $\Delta\Phi \in [0,\pi]$. Upper: $D$-$\bar D$ (a) and $B$-$\bar B$ (b) in $p$+$p$ collisions, the latter reconstructed through $b\to J/\psi$ decays. The calculation is at $\sqrt s=7$~TeV, and the $B$-$\bar B$ data combine the 7 and 8~TeV samples. Lower: $D$-$\bar D$ (c) and $B$-$\bar B$ (d) in $p$+Pb collisions at $\sqrt{s_{\rm NN}}=8.16$~TeV, the latter a prediction. Dashed lines include the perturbative and nonperturbative Sudakov factors ($S_{\rm P}+S_{\rm NP}$) or only the nonperturbative one ($S_{\rm NP}$), which is the $G\to0$ limit of the matching. The bands show the matched result for
$c=0.5$, $1.0$ and $2.0$, with $c=1.0$ shown as a solid line. LHCb data from~\cite{LHCb:2012aiv,LHCb:2017bvf,LHCb:2020jse}.}
    \label{fig:DD_BB_data_full}
\end{figure*}

In the self-normalized correlations of Fig.~\ref{fig:DD_BB_data_full}, the calculation with only the nonperturbative Sudakov factor (black dashed lines) gives an away-side peak about twice as high as the measured one for $D$-$\bar D$ pairs in $p$+$p$ collisions, which shows that the resummed soft-gluon radiation is the main source of the away-side broadening. Near the back-to-back limit, where $G\to1$ (see Fig.~\ref{fig:Q2k2ratio}), the matched result agrees with the one including both Sudakov factors (red dashed lines) to better than $5\%$ for all values of $c$, while on the near side it approaches the nonperturbative result. On the near side, $\Delta\Phi\to0$, the pair originates from a nearly collinear $g\to Q\bar Q$ splitting, whose propagator $1/(\Pt^2+m^2)$ is regulated by the quark mass. This mass suppression of collinear splitting, analogous to the dead cone, is much stronger for bottom than for charm quarks, so that the near-side peak is as higher than the away-side one for $D$-$\bar D$ pairs but about seven times lower for $B$-$\bar B$ pairs, a hierarchy also seen in the data. 

While the matched calculation describes the $B$-$\bar B$ correlation in $p$+$p$ collisions and the $D$-$\bar D$ correlation in $p$+Pb collisions within the experimental uncertainties over most of the angular range, for $D$-$\bar D$ pairs in $p$+$p$ collisions it describes the data between $\Delta\Phi\simeq\pi/3$ and $2\pi/3$ but lies above the away-side peak and below the near-side one. Since near $\Delta\Phi\simeq\pi$ the matched result is insensitive to $c$, this difference may point to contributions beyond leading order, in particular three-parton final states such as the splitting of a final-state gluon into a nearly collinear $c\bar c$ pair, which populate the near-side and intermediate regions and, after self-normalization, lower the away-side peak, with a smaller effect for bottom quarks.

Double parton scattering (DPS), not included in our calculation, is enhanced in $p$+Pb relative to $p$+$p$ collisions, since the two hard scatterings can take place on different nucleons~\cite{Strikman:2001gz,dEnterria:2012jam,Blok:2019fgg,Helenius:2019uge}, as confirmed by the about three times larger ratio of like-sign to opposite-sign charm pairs measured by LHCb~\cite{LHCb:2020jse}. These approximately uncorrelated pairs add a nearly flat pedestal, which may explain the data lying slightly above the calculation around $\Delta\Phi\simeq\pi/2$ in Fig.~\ref{fig:DD_BB_data_full}(c), and which must be subtracted before a comparison with our predictions.

Figure~\ref{fig:ratio} shows the $p$+Pb to $p$+$p$ ratios of the width and area of the away-side peak for two windows, $0.7\pi<\Delta\Phi<1.3\pi$, as in our light-hadron study~\cite{Caucal:2025zkl}, and $0.85\pi<\Delta\Phi<1.15\pi$. The area is the integral of the coincidence probability over the window and the width that of a Gaussian fitted on it, not quoted for $D$-$\bar D$ pairs at $1<p_T<2$~GeV, where the $p$+$p$ distribution is nearly flat. The area is suppressed by $13\%$ to $16\%$ in the wider window and by $14\%$ to $19\%$ in the narrower one, which is more sensitive to the depletion of the peak center by the broadening, with a weak dependence on flavor and $p_T$. Since the coincidence probability is normalized per trigger, this is the suppression of pairs relative to single mesons, whose yield probes the target at momenta of order $p_T$, while the away-side yield is dominated by imbalances $|\Kt|\ll p_T$, closer to $Q_{s,{\rm Pb}}$. For $D$-$\bar D$ pairs the width is enhanced by $23\%$ to $24\%$ at $2<p_T<4$~GeV, decreasing to $10\%$ to $11\%$ at $6<p_T<8$~GeV as the hard scale grows relative to $Q_{s,{\rm Pb}}$, in both windows. For $B$-$\bar B$ pairs it is enhanced by $11$ to $12\%$ in the wider and $14\%$ to $18\%$ in the narrower window, which resolves the sharper core of the $p$+$p$ peak (Supplemental Material~\ref{app:dihadron}). Varying $Q^2_{s0,{\rm Pb}}$ changes both ratios by $2\%$ to $5\%$ (up to $6\%$ for the width at $2<p_T<4$~GeV).

For forward di-$\pi^0$ correlations, in contrast, STAR observed a suppression of the away-side yield without significant broadening~\cite{STAR:2021fgw}, and our previous calculation predicted a width ratio close to unity at forward LHC kinematics~\cite{Caucal:2025zkl}, with the window of Fig.~\ref{fig:ratio}(a). The larger broadening for heavy mesons is a consequence of the heavy-quark mass. First, the mass reduces the collinear radiation, which broadens the peak in both $p$+$p$ and $p$+Pb collisions and therefore dilutes the nuclear broadening in the width ratio. In addition, the heavy meson carries a large fraction of the momentum of its parent quark (Fig.~\ref{fig:z1}), so that at fixed hadron $p_T$ the parent quark has a lower transverse momentum than for light hadrons and probes a denser region of the target.

\textbf{\textit{Summary and Outlook.}}
\label{sec:conclusions} 

We studied forward heavy-meson production and $D$-$\bar D$ and $B$-$\bar B$ azimuthal correlations in $p$+$p$ and $p$+Pb collisions at the LHC within the CGC effective theory at leading order, combining small-$x$ evolution with Sudakov resummation for massive quarks over the full azimuthal range. With one set of parameters, the calculation describes the LHCb single inclusive $D^0$ and $B^0$ spectra and the measured pair correlations. For $p$+Pb collisions at $\sqrt{s_{\rm NN}}=8.16$~TeV we predict a $13\%$ to $19\%$ suppression of the away-side yield and a broadening of its width by up to about $24\%$ for $D$-$\bar D$ pairs. While this broadening is small for forward light-hadron correlations, it is sizable for heavy mesons, as a consequence of the heavy-quark mass in both the radiation and the fragmentation of the heavy quarks.

Natural extensions of this work include next-to-leading order corrections and a comparison with a baseline calculation without saturation effects. We hope that this work will motivate measurements of the away-side width and yield in $p$+Pb collisions as a function of flavor and transverse momentum, which would provide a direct test of the interplay between Sudakov radiation and gluon saturation.

\textbf{\textit{Acknowledgements.}} We thank Paul Caucal, Zhong-Bo Kang, Piotr Korcyl, Tomasz Stebel, and Raju Venugopalan for various discussions and an earlier collaboration that motivated this work. F.S. thanks Qian-Shun Dai, Ming-Jun Liu, and Ding Yu Shao for helpful correspondence and for sharing their notes on heavy-quark pair production, which clarified the extraction of the double and single logarithms from the SCET factorization in Supplemental Material~\ref{app:sudakov}.  W.B.Z. thanks Cyrille Marquet, Yu Shi and Wenjing Xing for the fragmentation tables and helpful discussions. This work is supported by the U.S. Department of Energy, Office of Science, Office of Nuclear Physics, under DOE Contract No.~DE-SC0012704 (B.P.S.), and within the framework of the Saturated Glue (SURGE) Topical Theory Collaboration (F.S., B.P.S.). F.S. is supported by the Laboratory Directed Research and Development of Brookhaven National Laboratory and RIKEN-BNL Research Center. W. B. Z is supported by the Outstanding Leading Talent Team Program of Central China Normal University (XJ2026000302).

\bibliographystyle{utcaps}
\bibliography{refs}

\section{Supplemental Material}

In this Supplemental Material we first present the $D$-$\bar D$ and $B$-$\bar B$ coincidence probabilities in $p$+$p$ and
$p$+Pb collisions at $\sqrt{s_{\rm NN}}=8.16$~TeV, from which the ratios of Fig.~\ref{fig:ratio} of the manuscript are extracted, and the
average momentum fraction $\langle\xi\rangle$ of heavy mesons, which quantifies how closely the hadronic kinematics follows
the partonic one. We then examine the average matching weight as a function of $\Delta\Phi$, which shows where the
Sudakov-resummed result and the CGC result with the nonperturbative factor dominate. Finally, we discuss the structure
of the perturbative Sudakov factor for massive quarks and the parton-to-hadron kinematics.

\subsection{Dihadron correlations of $D$-$\bar D$ and $B$-$\bar B$ pairs}
\label{app:dihadron}
Figure~\ref{fig:DD_prediction} shows the $D$-$\bar D$ coincidence probability in the away-side region in the rapidity bin $2.7<y<3.7$ for several $p_T$ bins, computed with the parameters fixed in the main text, from which the ratios of Fig.~\ref{fig:ratio} are extracted. The width of the away-side peak results from two sources of transverse-momentum imbalance: soft-gluon radiation, described by the Sudakov factor, and the transverse momentum of the small-$x$ gluons of the target, of order $Q_s$. While the Sudakov factor is the same in $p$+$p$ and $p$+Pb collisions, the larger saturation scale of the nucleus broadens and suppresses the away-side peak. At $1<p_T<2$~GeV the quark momenta are comparable to the imbalance from Sudakov radiation and the target, so that the correlation has no well-defined away-side peak, and the nucleus suppresses the whole distribution by $13\%$ in the wider and $14\%$ in the narrower window, without a measurable broadening. Above $p_T=2$~GeV the width is enhanced by $24\%$, $13\%$ and $11\%$ in the three bins ($23\%$, $12\%$ and $10\%$ in the narrower window), decreasing with $p_T$ as $Q_{s,{\rm Pb}}$ becomes smaller relative to the hard scale, and the yield is suppressed by $13$ to $16\%$ ($14$ to $19\%$). Both yields approach the dilute limit with increasing $p_T$, the trigger yield faster than the away-side pairs, so that the per-trigger suppression increases slightly with $p_T$.

The Gaussian describes the calculated peak well: in $p$+$p$ collisions it deviates from it by at most $5\%$ ($6\%$) of its maximum for $D$-$\bar D$ ($B$-$\bar B$) pairs in the wider window and by at most $1\%$ ($2\%$) in the narrower one, and the area under it agrees with the integral of $CP(\Delta\Phi)$ to better than $0.3\%$ in the ratios. For $D$-$\bar D$ pairs the two windows give width ratios within $1\%$ of each other, while for $B$-$\bar B$ pairs the narrower window gives a larger broadening, as discussed below. The half width at half maximum, $\pi-\Delta\Phi_{1/2}$ with $CP(\Delta\Phi_{1/2})=CP(\pi)/2$, which requires no fit, gives width ratios of $1.20$ and $1.13$ for $D$-$\bar D$ pairs at $4<p_T<6$ and $6<p_T<8$~GeV (at $2<p_T<4$~GeV the $p$+$p$ distribution does not fall to half its maximum) and $1.19$ and $1.16$ for $B$-$\bar B$ pairs, so that the broadening does not rely on the Gaussian form, although its size depends on how the width is defined.

Figure~\ref{fig:BB_prediction} shows the corresponding $B$-$\bar B$ coincidence probabilities for $5<p_T<6$ and $6<p_T<8$~GeV, for which the peak in $p$+Pb collisions is suppressed by about $13\%$ ($15\%$ to $16\%$ in the narrower window) and broadened by $11\%$ to $12\%$ ($14\%$ to $18\%$). The narrower window is sensitive to the sharper core of the $p$+$p$ peak, with Gaussian widths in $p$+$p$ collisions of $\sigma^{pp}_{\Delta\Phi}=0.60$ and $0.54$~rad against $0.77$ and $0.66$~rad in the wider one. Since the nucleus adds nearly the same imbalance in both windows, about $0.3$ to $0.4$~rad in quadrature, the relative broadening is larger in the narrower window. While the nuclear broadening is comparable for the two flavors in the wider window, in the narrower one it is larger for $B$-$\bar B$ pairs, $14\%$ against $10\%$ at $6<p_T<8$~GeV.

Figure~\ref{fig:z1} shows the average momentum fraction $\langle\xi\rangle$ carried by $D$ and $B$ mesons for $p_{T1}=p_{T2}$ in $p$+Pb collisions. For $p_T>4$~GeV we find $\langle\xi\rangle\simeq0.64$ to $0.65$ for $D$ and $0.78$ to $0.79$ for $B$ mesons, decreasing slowly to $0.58$ and $0.74$ at $p_T=1$~GeV, compared with $\langle\xi\rangle\simeq0.4$ for forward light hadrons~\cite{Caucal:2025zkl}. The increase with the quark mass reflects that a fast heavy quark loses only a small fraction of its momentum when it picks up a light antiquark to form the meson~\cite{Peterson:1982ak}, so that at fixed hadron $p_T$ the parent quark carries $p_{T,Q}\simeq p_T/\langle\xi\rangle$, lower than for light hadrons.

\subsection{Kinematic range of the matching function}
\label{app:matching}

\begin{figure}
    \centering    
\includegraphics[width=0.9\columnwidth]{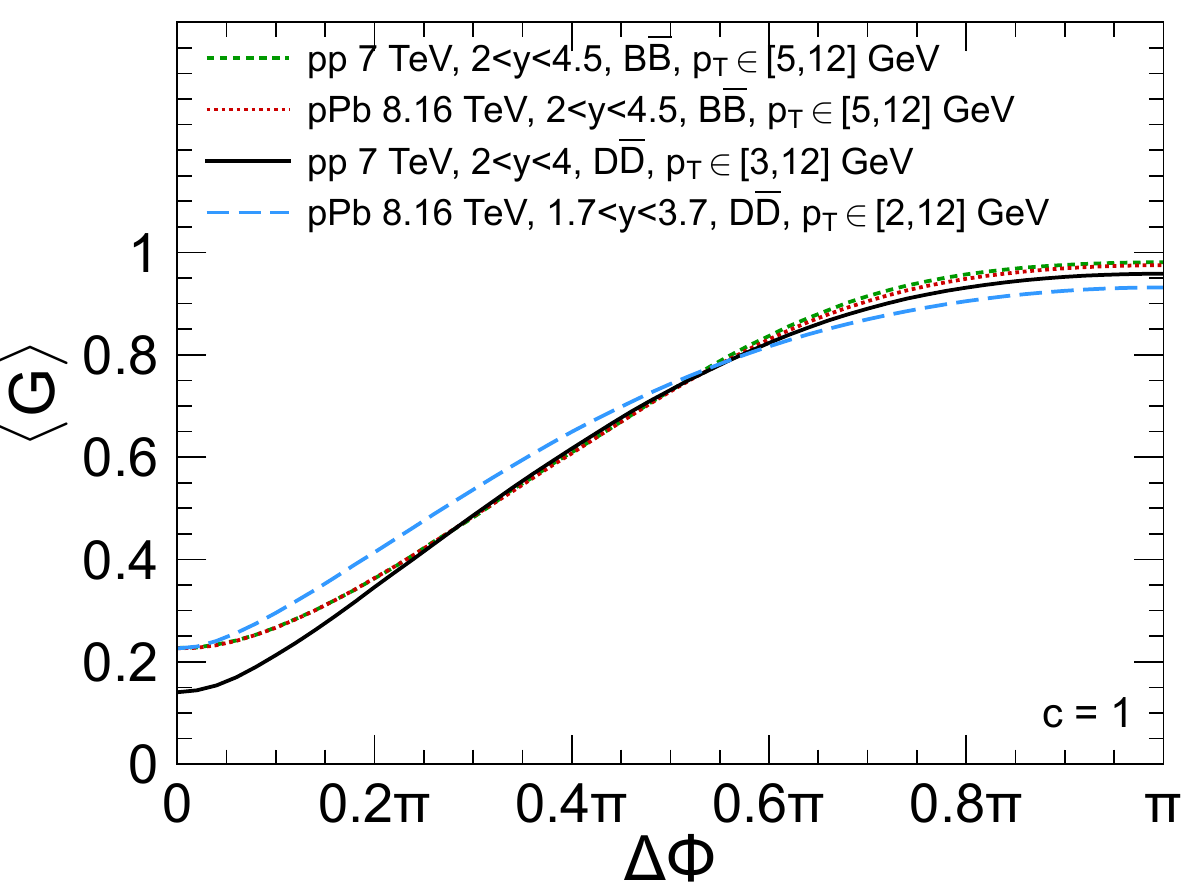}
    \caption{The average matching weight $\langle G\rangle$ for $c=1$ as a function of $\Delta\Phi$, weighted by the matched cross
section, for $D$-$\bar D$ and $B$-$\bar B$ pairs in $p$+$p$ collisions at $\sqrt{s}=7$~TeV and in $p$+Pb collisions at
$\sqrt{s_{\mathrm{NN}}}=8.16$~TeV, in the kinematics of Fig.~\ref{fig:DD_BB_data_full}.}
    \label{fig:Q2k2ratio}
\end{figure}

The matching function in Eq.~\eqref{eq:matching} depends on the ratio $R=M^2/\Kt'^2$ between the invariant mass of the heavy-quark pair and the momentum transferred by the target. It interpolates between the resummed result, $G\to1$ for $R\gg c^2$, and the CGC result with the nonperturbative Sudakov factor, $G\to0$ for $R\ll c^2$, with $G=1/2$ at $R=c^2$. To show which regions of $\Delta\Phi$ correspond to these two regimes, Fig.~\ref{fig:Q2k2ratio} displays the average weight $\langle G\rangle$ as a function of $\Delta\Phi$ for the four configurations of Fig.~\ref{fig:DD_BB_data_full}.

The behavior of $R$ can be understood from a simple partonic estimate, in which $\Kt'=\Kt$ at leading order. For a symmetric configuration with $|\ktone|=|\kttwo|=k$ and $y_1=y_2$, so that $z=1/2$, and relative azimuthal angle $\Delta\phi$ between the heavy quarks, one has $\Kt^2=4k^2\cos^2(\Delta\phi/2)$, $\Pt^2=k^2\sin^2(\Delta\phi/2)$ and $M^2=4(\Pt^2+m^2)$, so that
\begin{equation}
R=\frac{M^2}{\Kt^2}=\tan^2\frac{\Delta\phi}{2}+\frac{m^2}{k^2}\,\frac{1}{\cos^2(\Delta\phi/2)}\,.
\label{eq:R-estimate}
\end{equation}
Near the back-to-back configuration, $\Delta\phi\to\pi$, the imbalance vanishes and $R$ grows without bound, so that the Sudakov-resummed result dominates, as required by TMD factorization. Near the collinear configuration, $\Delta\phi\to0$, $R\to m^2/k^2$, which is below unity for the kinematics considered here, so that $G$ is small and the result with the nonperturbative Sudakov factor alone is approached. For $m\ll k$ the transition point $R=c^2$ is located at $\Delta\phi_c=2\arctan c$, that is $0.30\pi$, $0.50\pi$ and $0.70\pi$ for $c=0.5$, $1.0$ and $2.0$. Including the mass, $\sin^2(\Delta\phi_c/2)=(c^2-m^2/k^2)/(1+c^2)$, so that a finite quark mass shifts the transition to smaller $\Delta\phi$.

The full result in Fig.~\ref{fig:Q2k2ratio}, averaged over the momentum and rapidity bins and over the momentum transferred by the target, rises from $0.14$ to $0.23$ at $\Delta\Phi=0$ to $0.93$ to $0.98$ at $\Delta\Phi=\pi$, and crosses $1/2$ at $\Delta\Phi\simeq0.3\pi$. Since $\langle G\rangle$ exceeds $0.9$ for $\Delta\Phi\gtrsim0.7\pi$ ($0.8\pi$ for $D$-$\bar D$ pairs in $p$+Pb collisions), the away-side peak is dominated by the resummed result, and varying $c$ between $0.5$ and $2$ changes the width and area ratios of Fig.~\ref{fig:ratio} by at most $1\%$ in both windows. On the near side the perturbative Sudakov factor is suppressed, since the large momentum imbalance is generated by the hard $g\to Q\bar Q$ splitting and by multiple scattering rather than by soft-collinear radiation.

\subsection{Sudakov double and single logarithms for massive quarks}
\label{app:sudakov}
The perturbative Sudakov factor in Eq.~\eqref{eq:sudakov} collects the double and the leading single logarithms of $\mu^2 b^2$ generated by the incoming gluons and by the outgoing heavy-quark pair. The Sudakov logarithms associated with the massless incoming partons are those of the massless case~\cite{Mueller:2013wwa,Sun:2014gfa}: the collinear gluon from the proton and the small-$x$ gluon from the target each contribute a double logarithm with coefficient $C_A$, while the single logarithm $-2\beta_0$ accompanies only the collinear gluon distribution evaluated at $\mu_b$, as in our light-hadron study~\cite{Caucal:2025zkl}. We write $S_{\rm P}=S_{\rm P}^{g}+S_{\rm P}^{Q\bar Q}$, where $S_{\rm P}^{g}$ contains the $C_A$ and $\beta_0$ terms of Eq.~\eqref{eq:sudakov}, from the incoming gluons, and $S_{\rm P}^{Q\bar Q}$ the $C_F$ terms, from the final-state heavy-quark pair, which we extract here from the factorization for massive quarks~\cite{delCastillo:2020omr,vonKuk:2023jfd}.

For massless quarks the heavy-quark part reads
\begin{align}
    S_{\rm P}^{Q\bar Q}\Big|_{m=0}=\int_{\mu_b^2}^{\mu^2}\frac{\der\bar\mu^2}{\bar\mu^2}\,\frac{\alpha_s(\bar\mu)}{2\pi}\,2C_F\left[\ln\frac{\mu^2}{\bar\mu^2}-\frac{3}{2}\right],
    \label{eq:SP-massless}
\end{align}
which has the standard form $\int(\der\bar\mu^2/\bar\mu^2)\,[A\ln(\mu^2/\bar\mu^2)+B]$, where $A$ is the coefficient of the double logarithm and $B$ that of the single logarithm. At one loop $A=A^{(1)}\alpha_s/\pi$ and $B=B^{(1)}\alpha_s/\pi$, with values $A^{(1)}=C_F$ and $B^{(1)}=-3C_F/2$ in the notation of~\cite{Collins:1984kg}.

For massive quarks we follow the factorization in~\cite{delCastillo:2020omr,vonKuk:2023jfd}, in which the hard function is evolved from $\mu_b$ to $\mu$ with the massless anomalous dimensions, while the matching coefficient onto boosted heavy-quark effective theory (bHQET)~\cite{Fleming:2007qr,Fleming:2007xt} is evolved from $\mu_b$ to $m$. Written in the measure of Eq.~\eqref{eq:sudakov}, the heavy-quark part becomes
\begin{align}
    S_{\rm P}^{Q\bar Q}={}&\int_{\mu_b^2}^{\mu^2}\frac{\der\bar\mu^2}{\bar\mu^2}\,\frac{\alpha_s(\bar\mu)}{2\pi}\,2C_F\left[\ln\frac{\mu^2}{\bar\mu^2}-\frac{3}{2}\right]\nonumber\\
    &+\int_{\mu_b^2}^{m^2}\frac{\der\bar\mu^2}{\bar\mu^2}\,\frac{\alpha_s(\bar\mu)}{2\pi}\,2C_F\left[\ln\frac{\bar\mu^2}{m^2}+\frac{1}{2}\right].
    \label{eq:SP-massive}
\end{align}
The first line is the massless result, Eq.~\eqref{eq:SP-massless}, whose coefficients are the one-loop cusp and non-cusp anomalous dimensions of the massless quark current~\cite{Korchemsky:1987wg,Becher:2010tm}. The second line follows from the one-loop matching coefficient of the heavy-quark TMD fragmentation function onto bHQET, $C_m = 1 + \frac{\alpha_s C_F}{4\pi}\left[\ln^2(\bar\mu^2/m^2) + \ln(\bar\mu^2/m^2) + 4 + \pi^2/6\right]$~\cite{Fleming:2007xt,Hoang:2015vua,vonKuk:2023jfd}, squared for the pair, and its constant is the non-cusp part of the anomalous dimension of the bHQET current~\cite{Fleming:2007xt}. Below the mass the two logarithms add to $\ln(\mu^2/m^2)$, so that the double logarithm freezes at the mass, while the constant changes from $-3/2$ to $-1$. Writing the second line with $\theta(m^2-\bar\mu^2)$ inside the first integral gives the $C_F$ terms of Eq.~\eqref{eq:sudakov}.

At fixed coupling and for $\mu_b<m<\mu$, Eq.~\eqref{eq:SP-massive} gives
\begin{align}
    S_{\rm P}^{Q\bar Q}=\frac{\alpha_s C_F}{\pi}\left[\frac{1}{2} L_\mu^2 + L_\mu L_b - \frac{3}{2} L_\mu - L_b\right],
    \label{eq:SfQ}
\end{align}
with $L_\mu=\ln(\mu^2/m^2)$ and $L_b=\ln(m^2/\mu_b^2)$. The first two terms are the double logarithm with the mass as a floor, the third is the collinear single logarithm $B^{(1)}$ cut off at $m$, and the last is a soft single logarithm that survives below the mass. For $L_b\to0$ Eq.~\eqref{eq:SfQ} reduces to the massless result, and at fixed $\mu$ and $\mu_b$ the mass therefore removes $\frac{\alpha_s C_F}{\pi}\left[\frac{1}{2} L_b^2 - \frac{1}{2} L_b\right]$, which is the quantitative form of the dead-cone effect for this observable.

A previous heavy-flavor study~\cite{Marquet:2025jdr} also freezes the final-state double logarithm at the mass, but keeps only the singular part of the splitting function and therefore works at double-logarithmic accuracy, without single logarithms. The two results differ by $-\frac{\alpha_s C_F}{\pi}\left[\frac{3}{2} L_\mu + L_b\right]$. The $L_\mu$ term is the $B^{(1)}$ term of the massless case, cut off at $m$. The $L_b$ term has no massless counterpart: it is a soft effect of the dead cone.

\subsection{Parton-to-hadron kinematics}
\label{app:kinematics}

While for massless partons the hadron simply carries the fraction $\xi_i$ of the quark four-momentum, the heavy-quark and hadron masses are not negligible at the transverse momenta considered here, and $p_{hi}=\xi_ip_i$ would put the hadron off its mass shell. Following the light-cone definition of fragmentation~\cite{Collins:2011zzd}, we assume that the hadron carries no transverse momentum relative to the quark and moves along the quark direction $\hat{\vec p}_i$, and we take $\xi_i$ to be the fraction of the large light-cone component $\bar n_i\cdot p_i=E_i+|\vec p_i|$, with the light-like vectors $n_i^\mu=(1,\hat{\vec p}_i)$ and $\bar n_i^\mu=(1,-\hat{\vec p}_i)$ defined with respect to the quark direction. We denote three-vectors with an arrow, while bold symbols denote transverse vectors. The mass-shell condition then fixes the hadron momentum,
\begin{equation}
    p_{hi}^\mu=\frac{\xi_i}{2}\,(\bar n_i\cdot p_i)\,n_i^\mu+\frac{m_h^2}{2\,\xi_i\,\bar n_i\cdot p_i}\,\bar n_i^\mu\,.
\end{equation}
Since collinearity of massive particles and the light-cone fraction along the quark direction are not invariant under longitudinal boosts, this definition requires a choice of frame. We use the nucleon-nucleon center-of-mass frame, in which all our kinematics is evaluated, including the rapidities of the LHCb measurements in $p$+Pb collisions. A definition that is invariant under longitudinal boosts, in which $\xi_i$ is the fraction of the light-cone component $E_i+p_{z,i}$ along the beam axis and $\kthi=\xi_i\kti$, has a negligible effect in our single inclusive spectra.

Inverting the relation above, with $E_{hi}=m_{T,hi}\cosh y_{hi}$, $p_{z,hi}=m_{T,hi}\sinh y_{hi}$ and $|\vec p_{hi}|^2=\kthi^2+p_{z,hi}^2$, gives a quark energy and momentum that are linear in those of the hadron,
\begin{align}
    |\vec p_i|&=\frac{(m_h^2-\xi_i^2m^2)\,E_{hi}+(m_h^2+\xi_i^2m^2)\,|\vec p_{hi}|}{2\,\xi_i\,m_h^2}\,,\nonumber\\
    E_i&=\frac{(m_h^2+\xi_i^2m^2)\,E_{hi}+(m_h^2-\xi_i^2m^2)\,|\vec p_{hi}|}{2\,\xi_i\,m_h^2}\,.
\end{align}
Since the hadron and quark momenta have the same direction, $\kti=\kthi\,|\vec p_i|/|\vec p_{hi}|$ and $y_i=\frac{1}{2}\ln[(E_i+p_{z,i})/(E_i-p_{z,i})]$, with $p_{z,i}=p_{z,hi}\,|\vec p_i|/|\vec p_{hi}|$. The map rescales the three-momentum along a fixed direction. With $\der^2\boldsymbol k\,\der y=\der^3p/E$ and $\der|\vec p_i|/\der|\vec p_{hi}|=E_i/E_{hi}$, the Jacobian is
\begin{equation}
    \mathcal J_i=\frac{E_{hi}}{E_i}\,\frac{|\vec p_i|^2}{|\vec p_{hi}|^2}\,\frac{\der|\vec p_i|}{\der|\vec p_{hi}|}=\frac{\kti^2}{\kthi^2}\,,
\end{equation}
which reduces to $1/\xi_i^2$ for massless quark and hadron.

\end{document}